\documentclass[aps,eqsecnum,twocolumn,prd,showpacs,nofootinbib]{revtex4-1}
\usepackage{graphicx}
\def\bold#1{\setbox0=\hbox{$#1$}%
     \kern-.025em\copy0\kern-\wd0
     \kern.05em\%\baselineskip=18ptemptcopy0\kern-\wd0
     \kern-.025em\raise.0433em\box0 }
\def\slash#1{\setbox0=\hbox{$#1$}#1\hskip-\wd0\dimen0=5pt\advance
         to\wd0{\hss\sl/\/\hss}}
\newcommand{\be}{\begin{equation}}
\newcommand{\ee}{\end{equation}}
\newcommand{\bea}{\begin{eqnarray}}
\newcommand{\eea}{\end{eqnarray}}

\newcommand{\spur}[1]{\not\! #1 \,}

\begin{document}
\begin{titlepage}
\addtolength{\jot}{10pt}

\flushright{BARI-TH/08-600  }

\title{\bf Radiative transitions of heavy quarkonium states}

\author{Fulvia De Fazio \\}

\affiliation{  Istituto Nazionale di Fisica Nucleare, Sezione di
Bari, Italy }

\begin{abstract}
We  study radiative decays of heavy $Q{\bar Q}$ states, both for
$Q=c$ and $Q=b$, using an effective Lagrangian approach which
exploits spin symmetry for such states. We use
 existing data on radiative quarkonium transitions
 to predict some  unmeasured decay rates.
 We also discuss how these modes can be useful to understand the structure of $X(3872)$.
 \end{abstract}

\vspace*{1cm} 

\maketitle
\end{titlepage}

\newpage
\section{Introduction}\label{sec:intro}

Heavy quarkonium Physics was born in 1974 with the discovery of
the $J/\psi$, the first observed bound state of a heavy quark and
a heavy antiquark. Since then, quarkonium spectra and decays have
been thoroughly studied by means of potential models, lattice QCD,
QCD sum rules and effective theories (for recent reviews see
\cite{review, Voloshin:2007dx}). In particular, the agreement of
the observed mass levels and potential model predictions was
considered as a success of the latter, at least until 2003, when a
 series of observations of new states started
enriching our knowledge of $c{\bar c}$ and $b{\bar b}$ states and
stimulating new investigations. Indeed,   several aspects of such
new states seem not to be reconciled with  predictions. Thus, we
have to face two possibilities: either the accuracy of the
theoretical approaches  has to be questioned, or the newly
observed states are not conventional $Q{\bar Q}$ quarkonia
\cite{Swanson:2006st, reviewnew}.

In order to discuss these topics, it is useful to adopt the usual
classification of $Q{\bar Q}$ states in terms of the radial
quantum number $n$, the orbital angular momentum $L$, the spin $s$
and the total angular momentum $J$. The  state  identified by
$n^{2s+1}L_J$  corresponds to a meson  with parity $P=(-1)^{L+1}$
and charge-conjugation $C=(-1)^{L+s}$. In analogy with potential
model terminology, states with $L=0$ are referred to as $S$ wave
states, those with $L=1$ as $P$ wave, $L=2$ as $D$ wave states,
and so on.

In Table \ref{states} we collect  quarkonium resonances
corresponding to $S$, $P$ and $D$ wave states with $n=1$, and $S$
and $P$ wave states with $n=2$, which are the subject of our
analysis. In this Table, we include the established charmonium and
bottomonium states, together with their masses and  widths, when
known \cite{PDG}. Other known charmonium states are $\psi(4040)$,
$\psi(4160)$ and $\psi(4415)$, usually identified with the states
$3^3S_1$, $2^3D_1$ and $4^3S_1$, respectively, and therefore are
not included in the Table. As for bottomonium, the established
states not included in the Table are $\Upsilon(3S)$,
$\Upsilon(4S)$, $\Upsilon(5S)$, as well as the meson
$\Upsilon(11020)$, which is likely to be $\Upsilon(6S)$.

\begin{table*}\label{states}
\caption{Masses and widths of $1S$, $1P$, $2S$, $2P$ and $1D$
quarkonium states, taken from \cite{PDG}. The state
$\chi_{c2}(2P)$ is often referred to as $Z(3930)$.}
\begin{tabular}{|l|l||l|l|l||l|l|l|} \hline $n^{2s+1}L_J$ &
$J^{PC}$ &Charm & mass (MeV)& width (MeV)& Beauty
 & mass (MeV)& width (MeV) \\ \hline \hline $1^1S_0$ & $0^{-+}$ &
$\eta_c(1S)$  & $2980.3 \pm 1.2$ & $26.7 \pm 3.0$ & $\eta_b(1S)$ &
 $9300 \pm 20 \pm 20$ &
\\
 $1^3S_1$ & $1^{--}$& $J/\psi(1S)$ & $3096.916 \pm 0.011$ & $(93.2 \pm 2.1)\times 10^{-3}$ &
 $\Upsilon(1S)$  & $9460.30 \pm 0.26$ & $(54.02 \pm 1.25) \times 10^{-3}$ \\
 \hline \hline
$1^3P_0$ & $0^{++}$ & $\chi_{c0}(1P)$ & $3414.75 \pm 0.31 $ &
$10.5 \pm 0.8 $ & $\chi_{b0}(1P)$ & $9859.44 \pm 0.42 \pm 0.31 $ &
\\  $1^3P_1$ & $1^{++}$ & $\chi_{c1}(1P)$ & $3510.66 \pm
0.07$ & $0.88 \pm 0.05$ & $\chi_{b1}(1P)$ & $9892.78 \pm 0.26 \pm
0.31$ & \\ $1^3P_2$ & $2^{++}$ & $\chi_{c2}(1P)$ & $3556.20 \pm
0.09$ & $1.95 \pm 0.13$ &$\chi_{b2}(1P)$ & $9912.21 \pm 0.26 \pm
0.31 $ & \\  $1^1P_1$ & $1^{+-}$ & $h_c(1P)$ & $3525.93 \pm 0.27 $
& $<1$ & $h_b(1P)$ & & \\ \hline \hline $2^1S_0$ & $0^{-+}$
& $\eta_c(2S)$  & $3637 \pm 4$ & $14 \pm 7$ & $\eta_b(2S)$ & & \\
$2^3S_1$ & $1^{--}$& $\psi(2S)$ & $3686.093 \pm 0.034 $ & $(286
\pm 16) \times 10^{-3}$ & $\Upsilon(2S)$  & $10023.26 \pm 0.31$ &
$(31.98 \pm 2.63) \times 10^{-3}$ \\ \hline \hline
 $2^3P_0$ & $0^{++}$ & $\chi_{c0}(2P)$ &  &  &$\chi_{b0}(2P)$
 & $10232.5 \pm 0.4 \pm 0.5$ & \\
$2^3P_1$ & $1^{++}$ & $\chi_{c1}(2P)$ &  &  &$\chi_{b1}(2P)$
 & $10255.46 \pm 0.22 \pm 0.50$ & \\
$2^3P_2$ & $2^{++}$ & $\chi_{c2}(2P)$ & $3929 \pm5 \pm 2$  & $29
\pm 10 \pm 2$ &$\chi_{b2}(2P)$
 & $10268.65 \pm 0.22 \pm 0.50 $ & \\
$2^1P_1$ & $1^{+-}$ & $h_c(2P)$ & & & $h_b(2P)$ & & \\
 \hline
 \hline
$1^3D_1$ & $1^{--}$& $\psi(1^3D_1)$ & $3.775.2 \pm 1.7$ & $27.6
\pm 1.0$ & $\Upsilon (1^3D_1)$ & & \\
$1^3D_2$ & $2^{--}$& $\psi(1^3D_2)$ & & &$\Upsilon (1^3D_2)$ &
$10161.1 \pm 0.6 \pm 1.6$ & \\
$1^3D_3$ & $3^{--}$& $\psi(1^3D_3)$ & & &$\Upsilon (1^3D_3)$ &
& \\
$1^1D_2$ & $2^{-+}$& $\eta_{c2}(1^1D_2)$ & & &$\eta_{b2} (1^1D_2)$
&
 & \\
\hline
\end{tabular}\end{table*}

States below the open flavour threshold ($D{\bar D}$ for
charmonium, $B{\bar B}$ for bottomonium) are narrow, as well as
those states above such threshold whose strong decays to open
flavour are  forbidden by spin-parity conservation. For these
states  important decay modes  are radiative transitions, which
can be conveniently studied according to a perturbative expansion
of the Hamiltonian inducing the decay. In this way, one recognizes
that the most important transitions are electric dipole
transitions (named E1) and magnetic dipole transitions (M1). In
the former case quark spins are not flipped and the transitions
have $\Delta L=\pm 1$, $\Delta s=0$, while in the latter  quark
spin is flipped and $\Delta L=0$.  In the framework of potential
models, these can be calculated in terms of the
 wave functions of the involved quarkonium states, the overlap of which is
different from zero only for states with the same radial quantum
number $n=n^\prime$. This result is modified by the inclusion of
relativistic corrections, which induce non-zero transitions among
states with $n \neq n^\prime$ \cite{McClary:1983xw,Moxhay:1983vu}.

Another framework in which the analogy of quarkonia with an almost
non relativistic system is exploited is non relativistic QCD
(NRQCD) \cite{Bodwin:1992ye,Bodwin:1994jh},  an effective theory
based upon an expansion in the powers of $v$, the  relative
velocity of $Q$ and $\bar Q$ in the bound state. Several
predictions have been derived through this approach, the various
quantities (production cross sections, decay widths, etc) being
written as sums of contributions of several operators ordered
according to the velocity scaling rules \cite{Brambilla:2004jw}.

From an experimental point of view, there are several
possibilities to access quarkonium states. In the case of charm,
apart from $p{\bar p}$ production, direct production happens at
$e^+ e^-$ machines. Examples are CLEO-c at the center of mass
energy of $\psi(2S)$ and BES. Radiative decays of $\psi(2S)$ allow
to reach other states which cannot be directly produced from $e^+
e^-$ annihilation due to conservation of spin-parity, such as the
$\chi_{cJ}$ states.  $B$ factories have also revealed an important
source of charmonia. In this environment, $c{\bar c}$ states can
be produced $i$) through initial state radiation (ISR) when after
the emission of a photon from the initial state the effective
center of mass energy is suitable for the production of
charmonium; $ii$) in the collision of two photons radiated by $e^+
e^-$; $iii$) in $B$ decays. As for bottomonium, the same
mechanisms hold in principle (except for production in $B$
decays), even though the two photon collision has never succeeded
until now to produce bottomonia.

Radiative decays of quarkonia will play a role at the LHC. For
example, $\chi_{cJ}$ radiative decays to $J/\psi$ will be
considered by the ALICE experiment as a source of $J/\psi$ to
probe $J/\psi$ suppression in central heavy ion collisions
\cite{Gonzalez:2008nc}.

Thanks to this rich scenario of experimental facilities, several
new quarkonium states have been recently discovered. Among these,
some have  found their proper collocation in the above
classification and are included in Table \ref{states}: These are
the charmonia $h_c$ \cite{Rosner:2005ry}, $\eta_c(2S)$
\cite{Choi:2002na}, $\chi_{c2}(2P)$ (initially denoted by
$Z(3930)$) \cite{Uehara:2005qd} states, and, in the beauty case,
the $\eta_b(1S)$ meson \cite{:2008vj}.

Other states are still awaiting for the right interpretation,
since not only their quantum numbers are not well established, but
even their $Q{\bar Q}$ structure is questioned. We do not discuss
all of them here, but  focus only on  the state $X(3872)$ to which
part of our  analysis is devoted. This resonance was discovered
  by Belle Collaboration  as a narrow $J/\psi \pi^+ \pi^-$
  mass peak in exclusive $B^- \to K^- J/\psi \pi^+ \pi^-$ decay
  \cite{Choi:2003ue}, and
  later on confirmed by    CDF
\cite{Acosta:2003zx}, D0 \cite{Abazov:2004kp} and BaBar
\cite{Aubert:2004ns}.  The analysis of the $\pi^+ \pi^-$ mass
distribution shows that the two pions are likely to originate from
a $\rho^0$ decay. The subsequent measurement \cite{Abe:2005ix}:
$\displaystyle{{\cal B}(X \to \pi^+ \pi^- \pi^0 J/\psi) \over
{\cal B}(X \to \pi^+ \pi^- J/\psi)}=1.0 \pm 0.4 \pm 0.3$, showing
evidence of G-parity (isospin) violation, has been considered as
the argument against the charmonium interpretation for $X$ and in
favour of other exotic interpretations, in particular the
molecular one \cite{molecule}. However, as pointed out in
  \cite{Suzuki:2005ha}, assuming that the
 three pion mode originates from the decay $X \to J/\psi \, \omega$,
 the experimental ratio reported above is mainly due to the
 kinematical suppression of the $J/\psi \, \omega$ mode, and mechanisms can be found to explain
the ratio of the decay amplitudes, leaving
  the $c{\bar c}$ option still open.
  Several decay modes have been identified which might help
  discriminating a possible molecular structure of $X$ from the
  $c{\bar c}$ one, namely, decays to $\chi_{cJ} \pi(\pi)$
  \cite{Dubynskiy:2007tj} and radiative decays to $D^0{\bar D}^0 \gamma,\, D^+D^- \gamma$
  \cite{Voloshin:2005rt},
  even though the role of the latter ones is controversial \cite{Colangelo:2007ph}.

  If $X(3872)$ is a charmonium state, its possible quantum numbers have
  been discussed in \cite{Barnes:2003vb}. Among these, considering
  that the observation of the mode $X(3872) \to J/\psi \gamma$ \cite{Abe:2005ix} allows to
  fix $C=+1$, the most likely ones are the states $1^1D_2$ and
  $2^3P_1$.

In the following, we  study radiative decays of heavy $Q{\bar Q}$
states, both for $Q=c$ and $Q=b$, using an effective Lagrangian
approach which exploits spin symmetry for heavy $Q{\bar Q}$ states
\cite{Casalbuoni:1992yd}. The advantage of this method is
represented by the possibility of describing radiative transitions
between states belonging to the same $nL$ multiplet to states
belonging to another $n^\prime L^\prime$ one in terms of a single
coupling constant $\delta^{nLn^\prime L^\prime}$, allowing to use
data on known transitions to predict the yet unobserved ones.
Unlike the heavy-light $Q{\bar q}$ mesons, in heavy quarkonia
there is no heavy flavour symmetry \cite{Thacker:1990bm} because
of the infrared divergences developed in diagrams with two static
quarks exchanging gluons. Such divergences can be cured taking
into account the heavy quark kinetic energy operator, which is
${\cal O}(1/m_Q)$ and breaks heavy quark flavour symmetry. Because
of this, in our approach it is not  possible to exploit data on
charmonium to obtain quantitative information on bottomonium or
viceversa. However, we shall see that at a qualitative level,
bottomonium system can help in understanding charmonium.

Our first purpose in this paper is to exploit
 existing data on radiative quarkonium decays
 (we  always refer  to the states in Table
 \ref{states})
 to predict unmeasured decay rates.
 A second purpose is  to get insights on the proper
identification of states whose identity is still controversial,
starting  from the analysis of their radiative transitions. In
particular, this will be done in the case of $X(3872)$.

Our study concerns both  charmonium, both  bottomonium; in the
latter case,  we shall focus on the newly observed $\eta_b$ meson,
the lowest lying pseudoscalar $b{\bar b}$ state, and  the elusive
$J^{PC}=1^{+-}$ $h_b$ state.

\section{Effective Lagrangian for radiative transitions of D, P and S wave states }\label{sec:lagr}

Hadrons containing heavy quarks can be conveniently studied in the
infinite heavy quark mass limit. It is well known that in such a
limit new symmetries show up for systems containing a single heavy
quark, i.e. heavy quark spin and flavour symmetries. The effective
theory obtained from QCD in the heavy quark (HQ) limit and
displaying such symmetries is the heavy quark effective theory
(HQET), within which several advances have been obtained in heavy
quark Physics \cite{hqet}. In particular, due to spin symmetry,
states which differ only for the orientation of the heavy quark
spin with respect to the light degrees of freedom total angular
momentum are expected to be degenerate in the HQ limit. Such
states can be collected in  multiplets being $4 \times 4$ Dirac
matrices which, due to flavour symmetry, can describe  charmed
 and beauty states.

Something similar can be done in the case of heavy quarkonia, with
the limitation that flavour symmetry can no more be applied, so
that each multiplet  describes states with a defined heavy quark
flavour. The generic expression for a multiplet with relative
orbital angular momentum $L$ of the $Q{\bar Q}$ pair reads:
\begin{widetext}
\bea  J^{\mu_1 \dots \mu_L}&=&{ 1+ \spur{v} \over 2}\Big(
H_{L+1}^{\mu_1 \dots \mu_L \alpha } \gamma_\alpha + {1 \over
\sqrt{L(L+1)}} \sum_{i=1}^{L} \epsilon^{\mu_i \alpha \beta \gamma}
v_\alpha \gamma_\beta H_{L \gamma}^{\mu_1 \dots \mu_{i-1}
\mu_{i+1} \dots \mu_L}\nonumber \\ &+& {1 \over L} \sqrt{2L-1
\over 2L+1} \sum_{i=1}^{L}(\gamma^{\mu_i} -v^{\mu_i})
H_{L-1}^{\mu_1 \dots \mu_{i-1} \mu_{i+1} \dots
\mu_L}\label{multiplet} \\& -&{2 \over L \sqrt{(2L-1)(2L+1)}}
\sum_{i<j} (g^{\mu_i \mu_j}-v^{\mu_i}v^{\mu_j}) \gamma_\alpha
H_{L-1}^{\alpha \mu_1 \dots \mu_{i-1} \mu_{i+1} \dots \mu_{j-1}
\mu_{j+1}\dots \mu_L} \nonumber \\ &+&K_L^{\mu_1 \dots \mu_L}
\gamma_5 \Big){ 1- \spur{v} \over 2} \nonumber \eea
\end{widetext}
\noindent where  $v^\mu$ is  the heavy quark four-velocity and
$H_A$, $K_A$ are the effective fields of the various members of
the multiplets with total spin $J=A$. Since we  consider in the
following $S$, $P$ and $D$ wave states,  it is convenient to write
the corresponding multiplets   obtained from (\ref{multiplet}):
\begin{itemize}
\item L=2 multiplet: \bea  J^{\mu \nu}&=&{ 1+ \spur{v} \over
2}\Big\{ H_3^{\mu \nu \alpha } \gamma_\alpha \nonumber
 \\ &+& {1 \over \sqrt{6}}
 \left(\epsilon^{\mu \alpha \beta \gamma} v_\alpha
\gamma_\beta H_{2 \gamma}^\nu + \epsilon^{\nu \alpha \beta \gamma}
v_\alpha \gamma_\beta H_{2 \gamma}^\mu \right)\nonumber
\\&+& {1 \over 2} \sqrt{3 \over 5}
\left[(\gamma^{\mu} -v^{\mu}) H_1^\nu  + (\gamma^{\nu} -v^{\nu})
H_1^\mu \right]\label{Dwave} \\& -&{1 \over \sqrt{15}} (g^{\mu
\nu}-v^{\mu}v^{\nu}) \gamma_\alpha H_1^{\alpha} + K_2^{\mu
\nu}\gamma_5 \Big\}{ 1- \spur{v} \over 2} \,\,;\nonumber \eea
\item L=1 multiplet: \bea  J^{\mu }&=&{ 1+ \spur{v} \over 2}\Big\{
H_2^{\mu \alpha } \gamma_\alpha  + {1 \over \sqrt{2}}
\epsilon^{\mu \alpha \beta \gamma} v_\alpha \gamma_\beta H_{1
\gamma}\nonumber
\\&+& {1 \over \sqrt{3}}
(\gamma^{\mu} -v^{\mu}) H_0  + K_1^{\mu }\gamma_5 \Big\}{ 1-
\spur{v} \over 2}\,\,; \label{Pwave} \eea \item L=0 multiplet: \be
J={ 1+ \spur{v} \over 2} \left[H_1^\mu \gamma_\mu -H_0 \gamma_5
\right]{ 1- \spur{v} \over 2} \,\,.\label{Swave} \ee
\end{itemize}

Interactions of $Q{\bar Q}$ states can be described by  effective
Lagrangians written in terms of the effective fields $H$ and $K$
(for a review see \cite{Casalbuoni:1996pg}). This can be done for
the strong decays with emission of a light  meson and for the
radiative decays of interest here. One constructs effective
Lagrangians imposing Lorentz invariance, as well as invariance
under parity, charge conjugation and heavy quark spin symmetry
transformations. The corresponding transformations of the
multiplets are: \be J^{\mu_1 \dots \mu_L}
\stackrel{P}{\rightarrow}\gamma^0 J_{\mu_1 \dots \mu_L} \gamma^0
\label{parity} \ee \be J^{\mu_1 \dots \mu_L}
\stackrel{C}{\rightarrow}(-1)^{L+1} C [J^{\mu_1 \dots \mu_L}]^T C
\label{charge-conj} \ee \be J^{\mu_1 \dots \mu_L}
\stackrel{SU(2)_{S_h}}{\rightarrow}S J^{\mu_1 \dots \mu_L}
S^{\prime \dagger} \,\,.\label{HQspin} \ee In (\ref{HQspin}), $S$,
$S^\prime\in SU(2)_{S_h}$, $SU(2)_{S_h}$ being the group of heavy
quark spin rotations, with the property: $[S,{\spur
v}]=[S^\prime,{\spur v}]=0$.

 The effective Lagrangian describing
radiative transitions among members of the $P$ wave  and of the
$S$ wave multiplets has been derived in \cite{Casalbuoni:1992yd}:
\be {\cal L}_{nP \leftrightarrow mS}=\delta^{nPmS}_Q Tr
\left[{\bar J}(mS) J_\mu(nP) \right] v_\nu F^{\mu \nu} + \rm{h.c.}
\,,\label{lagPS} \ee where $\delta^{nPmS}_Q$ ($Q=c,b$) is a
coupling constant and $F^{\mu \nu}$ the electromagnetic field
strength tensor. The validity of the description is that of the
soft exchange approximation regime, when quarks are supposed to
exchange gluons of limited momenta. This is expected to work for
quarks of mass up to 80 GeV, as discussed  in
\cite{Casalbuoni:1992yd}.

Eq. (\ref{lagPS}) shows that  a single constant $\delta^{nPmS}_Q$
describes all the transitions among the members of the $nP$
multiplet and those of the $mS$ one. Indeed, the following decay
widths stem from (\ref{lagPS}) \cite{Casalbuoni:1992yd}: \bea
\Gamma(n^3P_J \to m^3S_1 \gamma) &=& {(\delta_Q^{nPmS})^2 \over
3 \pi} k_\gamma^3{M_{S_1} \over M_{P_J}}   \nonumber \\
\Gamma(m^3S_1 \to n^3P_J \gamma) &=& (2J+1) {(\delta_Q^{nPmS})^2
\over 9 \pi} k_\gamma^3{M_{P_J} \over M_{S_1}} \nonumber \\
\Gamma(n^1P_1 \to m^1S_0 \gamma) &=& {(\delta_Q^{nPmS})^2 \over 3
\pi} k_\gamma^3{M_{S_0} \over M_{P_1}}\label{E1widths}\\
\Gamma(m^1S_0 \to n^1P_1 \gamma) &=& {(\delta_Q^{nPmS})^2 \over
\pi} k_\gamma^3{M_{P_1} \over M_{S_0}} \nonumber
 \,,\eea where $k_\gamma$ is
the photon energy.

 Following the same guidelines leading to Eq. (\ref{lagPS}), we can construct the
effective Lagrangian describing  transitions among the members of
the $nD$  and  the $mP$ multiplets. Our result is: \be {\cal
L}_{nD \leftrightarrow mP}=\delta^{nDmP}_Q Tr \left[{\bar
J}_\alpha(mP) J_\mu^\alpha(nD) \right] v_\nu F^{\mu \nu} +
\rm{h.c.} \,,\label{lagDP} \ee which allows to compute the  decay
widths: \bea \hskip -0.6 cm\Gamma(m^3D_1 \to n^3P_0 \gamma)
&=&{5 \over 9} {(\delta_Q^{mDnP})^2
\over 3 \pi} k_\gamma^3{M_{P} \over M_{D}} \nonumber \\
\hskip -0.6 cm \Gamma(m^3D_1 \to n^3P_1 \gamma) &=& {5
\over 12} {(\delta_Q^{mDnP})^2
\over 3 \pi} k_\gamma^3{M_{P} \over M_{D}}  \label{D1Pwidths} \\
\hskip -0.6 cm \Gamma(m^3D_1 \to n^3P_2 \gamma) &=& {1
\over 36} {(\delta_Q^{mDnP})^2 \over 3 \pi} k_\gamma^3{M_{P} \over
M_{D}}  \,.\nonumber \eea
\noindent and
\bea \Gamma(m^1D_2 \to n^1P_1
\gamma) &=& {(\delta_Q^{mDnP})^2
\over 3 \pi} k_\gamma^3{M_{P} \over M_{D}} \nonumber \\
\Gamma(m^3D_2 \to n^3P_1 \gamma) &=& {(\delta_Q^{mDnP})^2
\over 4 \pi} k_\gamma^3{M_{P} \over M_{D}} \label{D2Pwidths}  \\
\Gamma(m^3D_2 \to n^3P_2 \gamma) &=& {(\delta_Q^{mDnP})^2 \over 12
\pi} k_\gamma^3{M_{P} \over M_{D}} \,, \nonumber \eea \noindent in
terms of a single new coupling constant $\delta^{mDnP}_Q$. We do
not consider the decays of the $^3D_3$ state, which proceed in $D$
wave and therefore are not described by the Lagrangian
(\ref{lagDP}).

 In the
following, we  make use of these results to study radiative
transitions of charmonia and bottomonia.

\section{Radiative transitions of  P and S wave states }\label{PS}

We  exploit the above results to systematically analyse some
radiative transitions among the states appearing in Table
\ref{states}. Some of the predictions stemming from the Lagrangian
(\ref{lagPS}) have been already obtained in
\cite{Casalbuoni:1992yd}, in such cases  we have exploited new or
more recent data.

\subsection{$1P \to 1S$ transitions}

The widths of the decay modes $\chi_{cJ}(1P) \to J/\psi \, \gamma$
can be obtained from the first equation in (\ref{E1widths}).
Experimental data are available for the three modes and  the
accuracy of spin-symmetry, which predicts that the transitions
$\chi_{cJ}(1P) \to J/\psi \, \gamma$ are all governed by the same
coupling constant $\delta_c^{1P1S}$, can be tested. Using
 \cite{PDG}: \bea {\cal B}(\chi_{c0}(1P) \to J/\psi \, \gamma)
&=& (1.28 \pm 0.11) \times 10^{-2}
 \nonumber \\
{\cal B}(\chi_{c1}(1P) \to J/\psi \, \gamma) &=& (36.0 \pm 1.9)
\times 10^{-2}
\label{BR1chicJ} \\
{\cal B}(\chi_{c2}(1P) \to J/\psi \, \gamma) &=& (20.0 \pm 1.0)
\times 10^{-2} \nonumber \eea together with the  $\chi_{cJ}$ full
widths  in Table \ref{states},  we obtain: $\delta_c^{1P1S}=0.227
\pm 0.013$ GeV$^{-1}$, $\delta_c^{1P1S}=0.241 \pm 0.009$
GeV$^{-1}$ and $\delta_c^{1P1S}=0.233 \pm 0.010$ GeV$^{-1}$,
respectively. Therefore, it is  correct to describe all these
modes in terms of a single constant,    the average value of which
is: \be \delta_c^{1P1S}=0.235 \pm 0.006 \,\,{\rm GeV}^{-1}
\label{deltac1P1S} \,.\ee

The same coupling $\delta_c^{1P1S}$ also governs the decay
$h_c(1P) \to \eta_c (1P) \,\gamma$, which has been observed but no
measurement of the rate has been determined, yet. Using the result
(\ref{deltac1P1S}) and the third equation in (\ref{E1widths}), we
predict: \be \Gamma(h_c(1P) \to \eta_c(1P) \, \gamma)=634 \pm 32
\, \rm{KeV}\,. \label{hctoetacwidth} \ee

The fact that the couplings $\delta_c^{1P1S}$ have been extracted
from the experimental data (\ref{BR1chicJ}) with values close to
each other supports the validity of the present approach in
exploiting the heavy quark spin symmetry to relate the $\chi_{cJ}
\to J/\psi \gamma$ ($J=0,1,2$) modes. Deviations from this result
would represent corrections to the heavy quark limit or, in the
language of the quark model, differences in the quarkonium state
wave functions. Actually, a quark model analysis supports the
present conclusions. In fact, the result (\ref{hctoetacwidth})
compares favorably with that obtained by Voloshin
\cite{Voloshin:2007dx}: $\Gamma(h_c(1P) \to \eta_c(1P) \,
\gamma)\simeq 0.65$ MeV, derived assuming the equality of the
radial wave function overlap integrals and exploiting the data on
$\chi_{cJ} \to J/\psi \gamma$ decays, a procedure similar to the
one adopted here. The outcome in (\ref{hctoetacwidth}) also agrees
with the result in \cite{Suzuki:2002sq}. Agreement is also met
with lattice QCD \cite{Dudek:2006ej}, whose predicted rate depends
on the use of lattice masses or of physical masses:
$\Gamma(h_c(1P) \to \eta_c(1P) \, \gamma)=663 \pm 132$ KeV or
$\Gamma(h_c(1P) \to \eta_c(1P) \, \gamma)=601 \pm 55$ KeV,
respectively. Notice that the estimated   uncertainty is sizeably
larger than in (\ref{hctoetacwidth}).

\begin{figure}[ht]
\begin{center}
\includegraphics[width=0.45\textwidth] {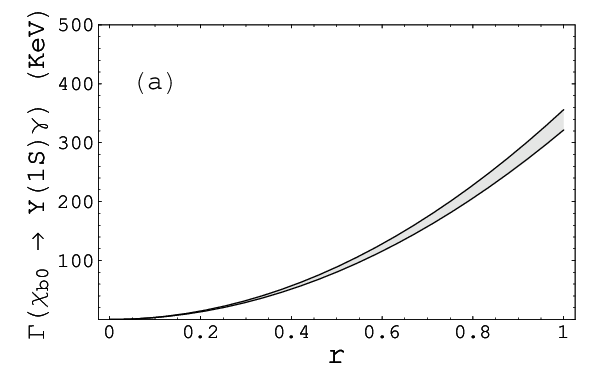} \hspace{0.5cm}
\includegraphics[width=0.45\textwidth] {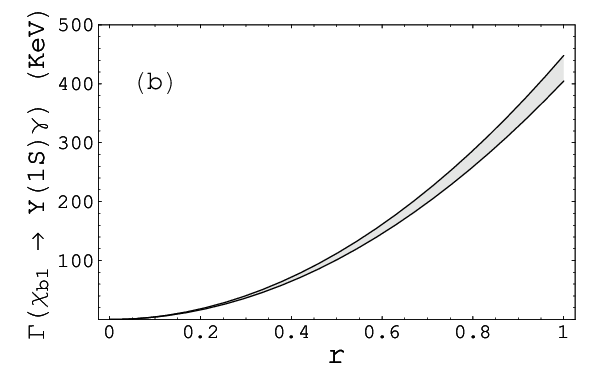}\\
\includegraphics[width=0.45\textwidth] {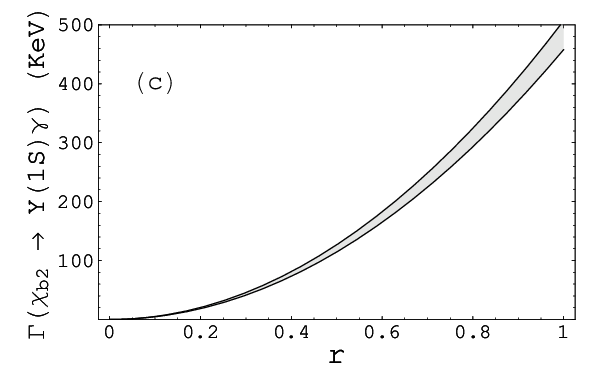}
\end{center}
\caption{\baselineskip=15pt  The widths $\Gamma(\chi_{b0} \to
\Upsilon \gamma)$ (KeV) (a), $\Gamma(\chi_{b1} \to \Upsilon
\gamma)$ (KeV) (b) and $\Gamma(\chi_{b2} \to \Upsilon \gamma)$
(KeV) (c), as a function of the ratio of effective couplings
$r=\delta_b^{1P1S}/\delta_c^{1P1S}$.} \vspace*{1.0cm}
\label{chibtoupsilon}
\end{figure}

For the corresponding beauty states $\chi_{bJ}(1P)$,  the
 available measurements \cite{PDG}: \bea {\cal
B}(\chi_{b0}(1P) \to \Upsilon(1S) \, \gamma) &<& 6 \times 10^{-2}
\nonumber \\
{\cal B}(\chi_{b1}(1P) \to \Upsilon(1S) \, \gamma) &=& (35 \pm 8)
\times 10^{-2}
 \label{BR1chibJ}\\
{\cal B}(\chi_{b2}(1P) \to \Upsilon(1S) \, \gamma) &=& (22 \pm 4)
\times 10^{-2} \nonumber  \eea  do not allow us to determine
$\delta_b^{1P1S}$ without a  measurement of the full width of a
$\chi_{bJ}(1P)$ state. Nevertheless, it is useful to study these
processes as  functions of the ratio
$r=\displaystyle{\delta_b^{1P1S} \over \delta_c^{1P1S} }$ of the
couplings, with the result  plotted in Fig. \ref{chibtoupsilon}.
We expect that the ratio $r$ is smaller than one, since it
includes the ratio of the beauty and the charm quark electric
charges: $\displaystyle{e_b \over e_c}$, as well as the effect of
the inverse heavy quark mass in each coupling $\delta$
\footnote{For example, in the case of $2S \to 2P$ transitions Eqs.
(\ref{deltac2S1P}), (\ref{deltab2S1P}) give $r \simeq 0.43$.}. As
a reference, we obtain that, at $r=0.5$, $\Gamma(\chi_{b0}(1P) \to
\Upsilon(1S) \, \gamma)=85 \pm 4$ KeV, $\Gamma(\chi_{b1}(1P) \to
\Upsilon(1S) \, \gamma)=107 \pm 5$ KeV and $\Gamma(\chi_{b2}(1P)
\to \Upsilon(1S) \, \gamma)=121 \pm 6$ KeV. Notice that, once the
value of $r$ in one decay mode has been determined, the prediction
for all the others follows.

\begin{figure}[ht]
\begin{center}
\includegraphics[width=0.45\textwidth]  {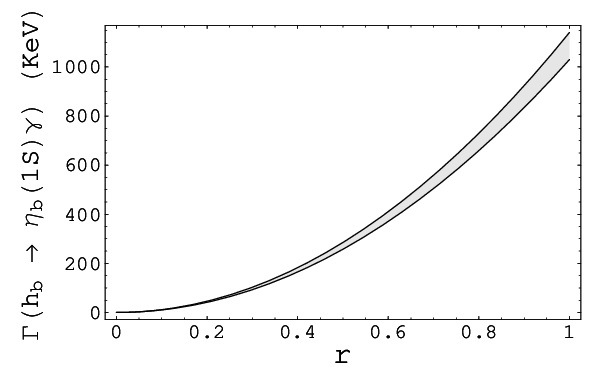}
\end{center}
\caption{\baselineskip=15pt  The width $\Gamma(h_b \to \eta_b
\gamma)$ (KeV)  as a function of the ratio of effective couplings
$r=\delta_b^{1P1S}/\delta_c^{1P1S}$.} \vspace*{1.0cm}
\label{hbtoetab}
\end{figure}
The same procedure can be applied to  the channel $h_b(1P)  \to
\eta_b(1S)  \, \gamma$, a mode to access the recently discovered
$\eta_b$ and to detect the still unseen $h_b$. We fix the $h_b$
mass   to the center of gravity of the $\chi_{bJ}$ states:
$M_{h_b}=\displaystyle{M_{\chi_{b0}}+3M_{\chi_{b1}}+5M_{\chi_{b2}}
\over 9}=9.89989$ GeV, an assumption supported by the
corresponding measurements in the charm sector; we obtain  the
result in Fig. \ref{hbtoetab}, which shows that for  $r= 0.5$ this
mode should have a width $\Gamma(h_b(1P)  \to \eta_b(1S)  \,
\gamma)=271 \pm 14$ KeV.

\subsection{$2S \to 1P$ transitions}

These transitions are described by the second equation in
(\ref{E1widths}). The measured branching fractions \cite{PDG}:
\bea {\cal B}(\psi(2S) \to \chi_{c0}(1P) \, \gamma) &=& (9.4 \pm
0.4) \times 10^{-2}
\nonumber \\
{\cal B}(\psi(2S) \to \chi_{c1}(1P) \, \gamma) &=& (8.8 \pm 0.4)
\times 10^{-2} \label{brpsi2S} \\
{\cal B}(\psi(2S) \to \chi_{c2}(1P) \, \gamma) &=& (8.3 \pm 0.4)
\times 10^{-2} \nonumber \eea together with  the measurement of
$\Gamma(\psi(2S))$ reported in Table \ref{states} permit to obtain
$\delta_c^{2S1P}=0.215 \pm 0.007 \,\, {\rm GeV}^{-1}$,
$\delta_c^{2S1P}=0.223 \pm 0.008 \,\, {\rm GeV}^{-1}$,
$\delta_c^{2S1P}=0.258 \pm 0.009 \,\, {\rm GeV}^{-1}$, and  the
average value: \be \delta_c^{2S1P}=0.228 \pm 0.005 \,\, {\rm
GeV}^{-1} \label{deltac2S1P} \,.\ee This value is close to that
obtained for $\delta_c^{1S1P}$, Eq. (\ref{deltac1P1S}), in analogy
to the outcome in \cite{Voloshin:2007dx} for the corresponding
radial wave function overlap integrals.

The result   (\ref{deltac2S1P}) allows us to predict the decay
width and the branching ratio of the mode $\eta_c(2S) \to h_c(1P)
\, \gamma$: \bea \Gamma(\eta_c(2S) \to h_c(1P) \, \gamma)&=& 21.1
\pm 0.9 \, \rm{KeV} \nonumber  \\ {\cal B}(\eta_c(2S) \to h_c(1P)
\, \gamma)&=& (0.15 \pm 0.08) \times 10^{-2}
.\label{etac2Stohcwidth} \eea This mode is interesting because it
represents another channel to study the  poorly known state
$h_c(1P)$. However, since the branching ratio turns out to be
tiny,  the observation is challenging.

\begin{figure}[ht]
\begin{center}
\includegraphics[width=0.45\textwidth] {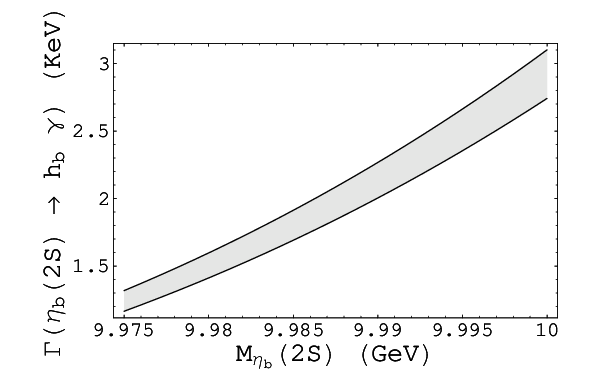} \\
\end{center}
\caption{\baselineskip=15pt  $\Gamma(\eta_b(2S) \to h_b \,
\gamma)$ (KeV) versus $M_{\eta_b(2S)}$ (GeV).} \vspace*{1.0cm}
\label{etab2Stohb}
\end{figure}

For the corresponding states with beauty, the following data are
available \cite{PDG}: \bea {\cal B}(\Upsilon(2S) \to \chi_{b0}(1P)
\, \gamma) &=& (3.8 \pm 0.4) \times 10^{-2}
\nonumber \\
{\cal B}(\Upsilon(2S) \to \chi_{b1}(1P) \, \gamma) &=& (6.9 \pm
0.4)
\times 10^{-2} \label{brUps2S} \\
{\cal B}(\Upsilon(2S) \to \chi_{b2}(1P) \, \gamma) &=& (7.15 \pm
0.35) \times 10^{-2} \,. \nonumber \eea Using the $\Upsilon(2S)$
 width in Table \ref{states}, we get: \be
\delta_b^{2S1P}=0.097 \pm 0.003 \,\, {\rm GeV}^{-1}  \,,
\label{deltab2S1P} \ee which can be used to predict the decay
width of the process $\eta_b(2S) \to h_b(1P) \, \gamma$. In Fig.
\ref{etab2Stohb} we show the result as a function of the unknown
mass of $\eta_b(2S)$, which we varied in a range obtained
considering the maximum and minimum value of the theoretical
determinations of the mass splitting $M_{\Upsilon(2S)}
-M_{\eta_b(2S)}$ \cite{Godfrey:2001eb}. Although the rate turns
out to be small, this mode can be considered as a possible channel
to detect $h_b$.

\subsection{$2P \to 1S$, $2S$ transitions}
As mentioned in the introduction,  Belle Collaboration has
observed a state,  $Z(3930)$, in $\gamma \gamma$ collision and
decaying to $D{\bar D}$, which can be most naturally identified
with $\chi_{c2}(2P)$ \cite{Uehara:2005qd}. For this state  no
radiative mode of the kind considered here has been detected, yet
\footnote{The decay to $\gamma \gamma$ has been observed.}. On the
other hand, radiative branching fractions of the $\chi_{bJ}(2P)$
to $\Upsilon(1S) \, \gamma$ and to $\Upsilon(2S) \, \gamma$ have
been measured \cite{PDG}, although this piece of information is
not enough  to determine the  couplings $\delta_b^{2P1S}$ and
$\delta_b^{2P2S}$ without  the measurement of the $\chi_{bJ}(2P)$
full widths. However, it is interesting to consider the  ratios:
\be R_J^{(b)}={\Gamma(\chi_{bJ}(2P) \to \Upsilon(2S) \, \gamma )
\over \Gamma(\chi_{bJ}(2P) \to \Upsilon(1S) \, \gamma )}
\label{ratios} \,.\ee We use \cite{PDG}: \bea {\cal
B}(\chi_{b0}(2P) \to \Upsilon(1S)
\, \gamma ) &=& (9 \pm 6) \times 10^{-3} \nonumber \\
{\cal B}(\chi_{b0}(2P) \to \Upsilon(2S) \, \gamma ) &=& (4.6 \pm
2.1)
\times 10^{-2} \nonumber \\
{\cal B}(\chi_{b1}(2P) \to \Upsilon(1S)
\, \gamma ) &=& (8.5 \pm 1.3) \times 10^{-2} \nonumber  \\
{\cal B}(\chi_{b1} (2P)\to \Upsilon(2S) \, \gamma ) &=& (21 \pm 4)
\times 10^{-2}  \label{chib2Pratios}\\
{\cal B}(\chi_{b2} (2P) \to \Upsilon(1S)
\, \gamma ) &=& (7.1 \pm 1.0) \times 10^{-2} \nonumber  \\
{\cal B}(\chi_{b2} (2P) \to \Upsilon(2S) \, \gamma ) &=& (16.2 \pm
2.4) \times 10^{-2} \,, \nonumber \eea which in turn provide: \be
R_0^{(b)}= 5.1 \pm 4.1 \,\,, \hskip 0.5cm R_1^{(b)}= 2.5 \pm 0.6
\,\,, \hskip 0.5cm R_2^{(b)}= 2.3 \pm 0.5 \,. \label{ratiosbexp}
\ee From these ratios we can extract the ratio of the coupling
constants $R_\delta^{(b)}=\displaystyle{\delta_b^{2P1S} \over
\delta_b^{2P2S}}$:
\par
\bea
  R_\delta^{(b)}&=&15 \pm 6 \nonumber \\
 R_\delta^{(b)}&=&9.3 \pm 1.1 \label{rdeltab}  \\
 R_\delta^{(b)}&=&8.4 \pm 0.9   \nonumber \eea
\noindent from $\chi_{b0}(2P)$, $\chi_{b1}(2P)$ and
$\chi_{b2}(2P)$ decays, respectively. These results show that also
in this case spin symmetry is fulfilled, even though in the case
of $\chi_{b0}(2P)$ the error affecting the result is  large. The
average value is: \be R_\delta^{(b)}=8.8 \pm 0.7 \,.
 \label{rdelta} \ee It is
reasonable that, even though the coupling might be different
passing from the beauty to the charm sector, the ratios of the
couplings  stay stable. Adopting such an assumption, one can
predict the corresponding ratios for $\chi_{cJ}(2P)$ states using
the result (\ref{rdelta}): \be R_2^{(c)}={\Gamma(\chi_{c2}(2P) \to
\psi(2S) \, \gamma ) \over \Gamma(\chi_{c2}(2P) \to \psi(1S) \,
\gamma )}=2.95 \pm 0.5 \label{ratioZ} \,.\ee This prediction can
be tested when new experimental data will be available and can be
used to support the identification of $Z(3930)$ with
$\chi_{c2}(2P)$.

 Interesting
considerations  stem for the case $J=1$. Actually, among the
canonical interpretations proposed for the puzzling state
$X(3872)$, a possible one is the identification with
$\chi_{c1}(2P)$. An important piece of experimental information
concerning $X(3872)$ is represented by the two measurements
\cite{:2008rn} \bea {\cal B}(B^+ \to XK^+, X\to
J/\psi\, \gamma ) &=& (2.8 \pm 0.8 \pm 0.2)\times10^{-6}  \nonumber\\
{\cal B}(B^+ \to XK^+, X\to \psi(2S) \, \gamma) &=& (9.9 \pm 2.9
\pm 0.6)\times 10^{-6}\,, \nonumber \\ \label{XBR} \eea from which
one has: \be R_X={\Gamma(X(3872) \to \psi(2S) \, \gamma ) \over
\Gamma(X(3872) \to \psi(1S) \, \gamma )}=3.5 \pm 1.4
\label{ratioX} \,.\ee If $X(3872)$ is identified as
$\chi_{c1}(2P)$, the above ratio $R_X$ can be computed  in our
framework, as done in (\ref{ratioZ}) in the case of $Z(3930)$. The
result is: \be R_1^{(c)}={\Gamma(\chi_{c1}(2P) \to \psi(2S) \,
\gamma ) \over \Gamma(\chi_{c1}(2P) \to \psi(1S) \, \gamma )}=1.64
\pm 0.25 \label{ratioXth} \,.\ee In view of the underlying
approximation, i.e. the equality of the ratio of the couplings in
the beauty and in the charm sector, we find that the experimental
value in (\ref{ratioX}) and the theoretical prediction obtained in
the hypothesis $X(3872)=\chi_{c1}(2P)$ are close enough to
consider this assumption plausible. This should be contrasted to
the composite scenarios, in which the mode  $X(3872) \to \psi(2S)
\, \gamma$ turns out to be suppressed compared to $X(3872) \to
\psi(1S) \, \gamma$ \cite{Swanson:2006st,Barnes:2003vb}.

\section{Radiative transitions of  D wave states }\label{D}
Experimental data on radiative transitions of $D$ wave states
exist in the case of $\psi(3770)$, usually identified with the
state $1^3D_1$ \footnote{We neglect  possible mixing with other
states.}. The following branching fractions are available
\cite{PDG}: \bea {\cal B}(\psi(3770) \to \chi_{c0}(1P) \gamma) &=&
(7.3 \pm 0.9)
\times 10^{-3}\nonumber \\
{\cal B}(\psi(3770) \to \chi_{c1}(1P) \gamma) &=&  (2.9 \pm 0.6)
\times 10^{-3} \,.\, \label{psi3770tochicJ} \eea From these data
we can extract the value of the coupling $\delta_c^{1D1P}$. The
average value obtained from the two modes above is: \be
\delta_c^{1D1P}=0.32 \pm 0.02 \,\, {\rm GeV}^{-1}
\,\,.\label{deltac1D1P} \ee In this case, a test of the validity
of spin symmetry is provided by the experimental analysis in
\cite{Briere:2006ff}, where the comparison of data with potential
model calculations supports the conjecture that the transition
matrix elements of $\psi(3770)$ to $\chi_{cJ}$ states are
independent on $J$.

The result (\ref{deltac1D1P}) allows us to predict width and
branching ratio of the third available radiative mode for
$\psi(3770)$: \bea \Gamma(\psi(3770) \to \chi_{c2}(1P) \, \gamma)
&=& 2.7 \pm 0.35
\,\, {\rm KeV}\,\, \nonumber\\
{\cal B}(\psi(3770) \to \chi_{c2}(1P)\, \gamma) &=&  (1.0 \pm 0.1)
\times 10^{-4} \,. \label{psi3770tochic2} \eea to be compared to
the experimental upper bound ${\cal B}(\psi(3770) \to
\chi_{c2}(1P)\, \gamma)<9 \times 10^{-4}$. For comparison, using
two variants of the potential model, Barnes et al.
\cite{Barnes:2005pb} find $\Gamma(\psi(3770) \to \chi_{c2}(1P)\,
\gamma)=3.3$ KeV or $\Gamma(\psi(3770) \to \chi_{c2}(1P)\,
\gamma)=4.9$ KeV, corresponding to the branching fraction  ${\cal
B}(\psi(3770) \to \chi_{c2}(1P)\, \gamma)=1.2 \times 10^{-4}$  or
${\cal B}(\psi(3770) \to \chi_{c2}(1P)\, \gamma)=1.8 \times
10^{-4}$, respectively, so that the prediction based on spin
symmetry is  different.

The same coupling governs all the transitions of the members of
the 1$D$ multiplet to the members of the 1$P$ one. Allowed decay
modes are:\bea
 1^3D_3 &\to& \chi_{c0,1,2}(1P) \gamma \nonumber \\
1^3D_2 &\to& \chi_{c1,c2}(1P) \, \gamma \nonumber \\ 1^1D_2 &\to&
h_c (1P)\, \gamma \,.\eea \noindent In the case of $1^3D_2$ the
decay to $\chi_{c0}$ is forbidden.

\begin{figure}[ht]
\begin{center}
\includegraphics[width=0.45\textwidth] {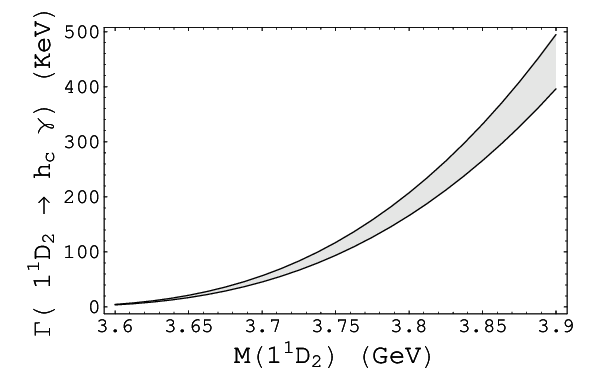} \\
\end{center}
\caption{\baselineskip=15pt  $\Gamma(1^1D_2 \to h_c \, \gamma)$
(KeV) versus the mass of the  $1^1D_2$ state  (in GeV).}
\vspace*{1.0cm} \label{1D2tohc}
\end{figure}

Analysing these modes is interesting not only {\it per se}, but
also in view of the already mentioned possibility that $X(3872)$
might be identified with the state $1^1D_2$. For this purpose, in
Fig. \ref{1D2tohc} we plot $\Gamma(1^1D_2 \to h_c \, \gamma)$
versus $M(1^1D_2) $. If $X$ coincides with the $1^1D_2$ state,
hence $M(1^1D_2)=3872$ MeV, we find that $\Gamma(1^1D_2 \to h_c \,
\gamma)=359 \pm 40$ KeV, while, if we use the masses reported in
\cite{Barnes:2005pb}, i.e. $M(1^1D_2)=3799 $ MeV or
$M(1^1D_2)=3837 $ MeV (depending on  the variant of the potential
model), we find: $\Gamma(1^1D_2 \to h_c \, \gamma)=185 \pm 20$ KeV
or $\Gamma(1^1D_2 \to h_c \, \gamma)=267 \pm 30$ KeV, to be
compared to the results in \cite{Barnes:2005pb}: $\Gamma(1^1D_2
\to h_c \, \gamma)= 339$ KeV or $\Gamma(1^1D_2 \to h_c \, \gamma)=
344$ KeV. In all cases, the decay width to $h_c$ is rather
sizeable.

\begin{figure}[ht]
\begin{center}
\includegraphics[width=0.45\textwidth] {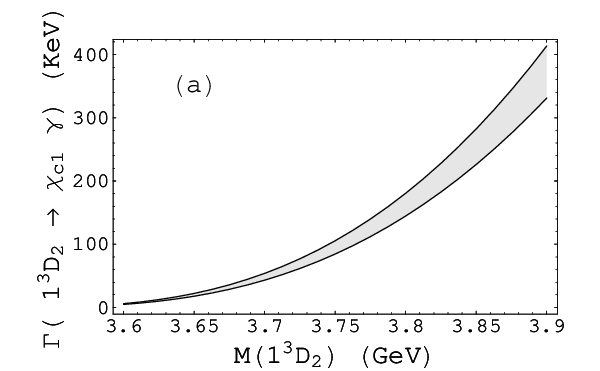} \hspace{0.5cm}
\includegraphics[width=0.45\textwidth] {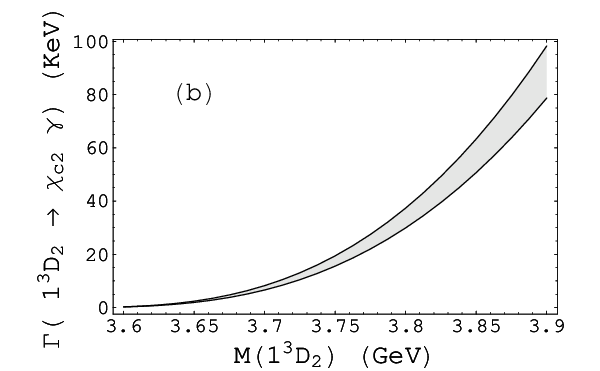}  \\
\end{center}
\caption{\baselineskip=15pt  Widths $\Gamma(1^3D_2 \to \chi_{c1}
\, \gamma)$ (KeV) (a) and $\Gamma(1^3D_2 \to \chi_{c2} \, \gamma)$
(KeV) (b) versus the mass of the $1^3D_2$ state (in GeV).}
\vspace*{1.0cm} \label{3D2tochic}
\end{figure}

The same  analysis can be carried out in the case of the state
$1^3D_2$, which was  initially proposed as a possible
identification for $X(3872)$, but is now  ruled out because of the
C-parity of this state (opposite to the one fixed for $X$). We
compute the decay widths to $\chi_{c1} \gamma$ and $\chi_{c2}
\gamma$ as a function of $M(1^3D_2)$, as plotted in Fig.
\ref{3D2tochic} and, in particular, in correspondence to the
masses reported in \cite{Barnes:2005pb}: $M(1^3D_2)=3800$ MeV or
$M(1^3D_2)=3838$ MeV, finding: $\Gamma(1^3D_2 \to \chi_{c1}\,
\gamma)=163 \pm 18$ KeV or $\Gamma(1^3D_2 \to \chi_{c1}\,
\gamma)=230 \pm 25$ KeV and $\Gamma(1^3D_2 \to \chi_{c2}\,
\gamma)=34 \pm 4$ KeV or $\Gamma(1^3D_2 \to \chi_{c2}\, \gamma)=51
\pm 6$ KeV. For comparison,  in \cite{Barnes:2005pb} the following
results are obtained: $\Gamma(1^3D_2 \to \chi_{c1}\, \gamma)=307$
KeV or $\Gamma(1^3D_2 \to \chi_{c1}\,  \gamma)=268$ KeV and
$\Gamma(1^3D_2 \to \chi_{c2}\,  \gamma)=64$ KeV or $\Gamma(1^3D_2
\to \chi_{c2}\,  \gamma)=66$ KeV.

\section{Conclusions }
We have analysed radiative decays of several $c{\bar c}$ and
$b{\bar b}$ states, using an effective Lagrangian approach valid
for heavy quarkonia. Exploiting existing data has allowed us to
derive model independent predictions on channels related by the
heavy quark spin symmetry. When available, experimental data are
consistent with the description based on this symmetry.

We have also considered the case of X(3872),  finding that the
 observed radiative modes are compatible with the
identification of this state with $\chi_{c1}(2P)$. As for its
identification with the $1^1D_2$ state, we have predicted the
decay rate of $1^1D_2 \to h_c \, \gamma$ as a function  of
$M(1^1D_2)$ and, in particular, for $M(1^1D_2)=3872$ MeV. The
observation of this decay for $X(3872)$ in agreement/disagreement
with such a prediction would support/discard this option. This
mode is anyway interesting, representing another channel to access
the state $h_c$,  one of the newly confirmed charmonium states.

In the beauty sector, we have considered some modes  involving the
$\eta_b$ and $h_b$ states, among which only the $\eta_b(1S)$ has
been recently discovered. We find that the mode $h_b \to \eta_b
\gamma$ could be detectable. As for the production of $h_b$ in
$\eta_b(2S)$ radiative decay, we predict a small rate.

\vspace*{1cm} \noindent {\bf Acnowledgments}  \par I thank P.
Colangelo and T.N. Pham for  reading this manuscript and for
discussions. This work was supported in part by the EU contract
No. MRTN-CT-2006-035482, "FLAVIAnet".

 
\end{document}